\def\Granadadep{Departamento de F\'\i sica Te\'orica y del Cosmos,
Facultad
de Ciencias, Universidad de Granada, Campus de Fuentenueva, Granada
18002, Spain}

\def\Granadainst{Instituto de F\'\i sica Te\'orica y Computacional
Carlos I, Facultad
de Ciencias, Universidad de Granada, Campus de Fuentenueva, Granada
18002, Spain}

\def\Valencia{IFIC, Centro Mixto Universidad de Valencia-CSIC,
Burjasot
              46100-Valencia,Spain.}
\def\Comision{Work partially supported by the DGICYT.}

 \def\w{\omega}
 \def\medio{\frac{1}{2}}
 \def\nn{\nonumber}
 \def\ni{\noindent}

 \def\cinf{$c\rightarrow\infty\;$}
 \def\PHO{{\cal P}^{HO}}
 
 \def\P{\hat{p}}
 \def\X{\hat{x}}
\def\PI{\hat{\pi}}
 \def\Q{\hat{q}}

\def\H{\hat{H}}
\def\Z{\hat{z}}
\def\Zc{\hat{z}^{\dag}}
\def\C{\hat{a}}
\def\Cc{\hat{a}^{\dag}}
\def\XLz{\tilde{X}^{L}_{z}}
\def\XLzc{\tilde{X}^{L}_{z^*}}

\documentstyle[12pt]{article}

\begin{document}

\setcounter{page}{0}

\begin{titlepage}



\begin{center}

{\bf{\LARGE CANONICAL COHERENT STATES FOR THE \\ \vskip 0.2cm
      RELATIVISTIC HARMONIC OSCILLATOR$^1$ }}
\end{center}

\bigskip
\bigskip

\centerline{ V. Aldaya$^{2,3}$
             and J. Guerrero$^{2,4}$ }     

\bigskip
\centerline{March 14, 1995}
\bigskip

\footnotetext[1]{\Comision}
\footnotetext[2]{\Granadainst} \footnotetext[3]{\Valencia}
\footnotetext[4]{\Granadadep}

\bigskip

\begin{center}
{\bf Abstract}
\end{center}

In this paper we construct manifestly covariant relativistic coherent
states on the entire complex plane which reproduce
others previously introduced on a given $SL(2,R)$ representation,
once a change
of variables $z\in C\rightarrow z_D \in $ unit disk is performed. We
also introduce
higher-order, relativistic creation and annihilation operators,
$\C,\Cc$, with canonical
commutation relation $[\C,\Cc]=1$ rather than the covariant one
$[\Z,\Zc]\approx$
Energy and naturally associated with the  $SL(2,R)$ group. The
canonical (relativistic) coherent
states are then defined as eigenstates of $\C$. Finally, we construct
a canonical,
minimal representation in configuration space by mean of eigenstates
of a
canonical position operator. \\

\vfil

\end{titlepage}

\vfil\pagebreak

\section{Introduction}
Ordinary coherent states were introduced from the beginning of the
developments of
Quantum Mechanics and Radiation Theory in several different, yet
equivalent, ways
according to different interesting properties with direct
applications to practical,
mainly optical, systems (see the pioneer work by Glauber
\cite{Glauber}). Essentially,
these states can
be characterized by a) giving minimal and symmetric $q-p$ uncertainty
relations, b) being
eigenstates of the annihilation operator $\C$, or c) as the result of
applying the
displacement operator $e^{i(\alpha \Cc-\alpha^* \C)}$ on the vacuum
$|0>$, and among
the practical properties, we mention the low noise in amplifying
applications
(as a consequence of a)) and optical coherence (as a consequence of
b)) (see for
instance \cite{Yuen,Dodonov}).

Relativistic quantum mechanical systems in general are characterized
by possessing
manifestly covariant commutation relations of the form
$[\hat{x},\hat{p}]\approx$
Energy, so that the uncertainty relations are no longer $\Delta
\hat{x}\Delta
\hat{p}\geq \frac{\hbar}{2}$, but  $\Delta \hat{x}\Delta \hat{p}\geq
\frac{\hbar}{2}|<\frac{\hat{E}}{mc^2}>|$ and, therefore, the absolute
minimum can be reached
only by the vacuum. In particular, the adopted commutation relations
for the basic
operators $\hat{x}, \hat{p}$ and $\hat{E}$ corresponding to the
quantum relativistic
harmonic oscillator are:
\begin{equation}
[\hat{E},\hat{x}]=-i\frac{\hbar}{m}\hat{p},\,\,\,\,\,\,
[\hat{E},\hat{p}]=im\omega^{2}\hbar\hat{x},\,\,\,\,\,\,
[\hat{x},\hat{p}]=i\hbar(1+\frac{1}{mc^{2}}\hat{E})\, ,
\label{Epxalg}
\end{equation}

\ni which implement a central (pseudo-)extension of the Lie algebra
(see Ref.
\cite{Jacobson} for a study of the cohomology of Lie algebras)
$SL(2,R)$ ($\approx$ Anti-deSitter group in 1+1 dimensions). This
algebra,
where $\hat{E}$ generates the time translations, $\hat{p}$ the space
translations and $\hat{x}$ the {\it boosts},
reproduces the pseudo-extended Poincar\'e algebra in the
$\omega\rightarrow 0$
limit and the extended Newton (non-relativistic harmonic oscillator)
algebra when $c\rightarrow\infty $.
It should be recalled at this point that it is the pseudo-extended
Poincar\'e group which regains the centrally extended
Galilei group in the non-relativistic limit \cite{Saletan} (for a
general study of
central extensions of groups see Ref. \cite{Bargmann}.

The deviation of the relativistic commutation relations between
$\hat{x}$ and $\hat{p}$
from the Galilean ones causes the different definitions of coherent
states given above
to be non-equivalent. The definition c) seems to be more widely
adopted at least for those
cases with an underlying group structure
\cite{Perelomov,Perelomovbook}).

In this paper we consider a set of states of the relativistic
harmonic oscillator
-our relativistic coherent states- obtained, in a natural way,
following a
group approach to quantization \cite{GAQ,anomalia},  which
 reproduce those previously introduced by
Perelomov \cite{Perelomov} after the change of variables
\begin{equation}
z_D=\sqrt{\frac{2}{N}}\frac{z}{1+\kappa}\,,\,\,\,\, N\equiv
\frac{mc^2}{\hbar \omega} \,, \,\,\,\kappa\equiv
\sqrt{1+\frac{2zz^*}{N}} ,
\end{equation}

\ni from $C$ to the unit disk $D$, has been performed. Then, using a
generalization
of the concept of Polarization in Geometric Quantization
\cite{Souriau,Kostant},
we are able to find
higher-order creation and annihilation operators $\C, \Cc$ in terms
of the basic
generators $\Z,\,\Zc$ of $SL(2,R)$, as a non-polynomic function,
satisfying
canonical
(yet relativistic) commutation relations, and allowing for a
conventional way of
defining canonical, relativistic coherent states as eigenstates of
the new
higher-order annihilation operator $\C$. The new relativistic
coherent states
 thus satisfy properties fully analogous to those of ordinary
(non-relativistic)
coherent states, although defined in terms of canonical or Darboux
\cite{Abraham} co-ordinates.

Our construction  of the operators $\C$ and $\Cc$ in terms of the
$SL(2,R)$ generators $\Z$ and $\Zc$  is invertible, but the inverse
(non-polynomical) relation, i.e. $\Z$ and $\Zc$ in terms of $\C$ and
$\Cc$
must not be confused, however, with the ``standard" quadratic
realization of
the
$SL(2,R)$ generators as $\hat{K}_-=\medio(\C)^2,\,\,
\hat{K}_+=\medio(\Cc)^2$
and $\hat{K}_3=\medio \Cc \C$, appearing, for instance, in Quantum
Optics.
This quadratic construction close algebra with the operators $\C$ and
$\Cc$
themselves but possesses the drawback that only two particular
irreducible
representations of $SL(2,R)$ can be obtained, namely those with
Bargmann index
$k=\frac{1}{4},\frac{3}{4}$.

\section{Relativistic coherent states (RCS).}

In the group-quantization scheme, the coherent states (generalizing
the
standard non-relativistic coherent states \cite{Glauber}), as well as
the
corresponding wave functions, are defined by mean of infinitesimal
relations
(differential polarization equations) \cite{SL(2R)}, rather than a
finite group action on the vacuum,
associated with a previously given representation of the group
\cite{Perelomov,Perelomovbook}
(see \cite{Klauder,Klauder-MacKenna,Klauder-Sudarshan} for a more
general study of overcomplete families of
states non-necessarily associated with groups). These are defined
simply as
\cite{SL(2R),RelBargmann}:
\begin{eqnarray}
|z> &\equiv & \sum_{n=0}^\infty \tilde{\Phi}^N_n(z,z^*)^* |N,n>
\,\,\,\leftrightarrow \,\,\,
     \tilde{\Phi}^N_n(z,z^*)=<z|N,n> \\
\tilde{\Phi}^N_n(z,z^*) &\equiv&
\frac{1}{\pi\sqrt{n!}}\sqrt{\frac{(2N)_n}
{(2N)^n}}\sqrt{\frac{2N-1}{2N}}
  \left(\frac{1+\kappa}{2}\right)^{-N-n}{z^{*}}^{n} \nn
\end{eqnarray}

\ni where the states $|N,n>$ constitute the Fock space for the
relativistic
harmonic oscillator, i. e.
\begin{equation}
 <0,N|N,0>=1,\;\;\; |N,n>=\frac{(\Zc)^{n}|N,0>}
{\sqrt{n!\frac{(2N)_n}{(2N)^n}}} \, ,
 \end{equation}
 \begin{eqnarray}
 \Z|N,n> &=& \sqrt{n(1+\frac{n-1}{2N})}|N,n-1> \nn \\
 \Zc|N,n> &=& \sqrt{(n+1)(1+\frac{n}{2N})}|N,n+1> \label{rep} \\
 \H|N,n> &=& n|N,n> \nn
 \end{eqnarray}

\ni They carry an irreducible representation (with Bargmann index
$k=N$)
of the $SL(2,R)$ algebra
\begin{equation}
\left[ \H,\Z \right] = -\Z \, , \,\,\,\,\,\;
\left[ \H,\Zc \right] = \Zc \, ,\,\,\,\,\,\;
\left[ \Z,\Zc \right] = \hat{1} +\frac{1}{N}\H \label{commut}
\end{equation}

\ni where these operators are essentially the right generators of the
$SL(2,R)$
group \cite{SL(2R),oscilata}. The relation with the standard notation
for
(abstract) $SL(2,R)$ generators is
$\Z=\frac{1}{\sqrt{2N}}\,\hat{K}_-=
\frac{1}{\sqrt{2N}}\,(\hat{K}_1-i\hat{K}_2),\,\,\,
\Zc=\frac{1}{\sqrt{2N}}\,\hat{K}_+=
\frac{1}{\sqrt{2N}}\,(\hat{K}_1+i\hat{K}_2)\,\,$
and $\H=\hat{K}_3-N$. However, our generators have
the advantage of a proper non-relativistic limit.

\ni The associated wave functions $<z|z'>$ are:
\begin{eqnarray}
<z'|z>&=&\sum^{\infty}_{n=0}
\tilde{\Phi}^N_n(z',z'{}^*)\tilde{\Phi}^N_n(z,z^*)^*\nn \\
&=&\frac{1}{\pi}\frac{2N-1}{2N}\left(\frac{1+\kappa}{2}\right)^{-N}
    \left(\frac{1+\kappa'}{2}\right)^{-N}
\sum^{\infty}_{n=0}\frac{(2N)_n}{n!(2N)^n}
   \left(\frac{2z'^*}{1+\kappa'}
   \frac{2z}{1+\kappa}\right)^n \\
  &=&\frac{1}{\pi}\frac{2N-1}{2N}\left(\frac{1+\kappa}{2}\right)^{-N}
\left(\frac{1+\kappa'}{2}\right)^{-N}
\left[1-\frac{2z'^*z}{N(1+\kappa')(1+\kappa)}\right]^{-2N} \nn
\end{eqnarray}

As in the non-relativistic case, the RCS constitute an overcomplete
set and
satisfy the reproducing kernel property with respect to the group
measure:
\begin{eqnarray}
 I&=&\int \frac{dzdz^*}{\kappa}|z><z| \\
 |z'>&=&\int \frac{dzdz^*}{\kappa}|z><z|z'>
\end{eqnarray}

The expectation values of $\Z$ and $\Zc$ on the coherent states are
$<\Z>\equiv \frac{<z|\Z|z>}{<z|z>} = z \,\,\,$ and $\,\,\,<\Zc> =
z^*,\,\,$
making the variables $z,z^* \in C$ especially suitable to describe
the
Bargmann-Fock-like representation. Defining the operators $\X$ and
$\P$ in the
usual way, i.e.
\begin{equation}
\X =\sqrt{\frac{\hbar}{2m\w}}\left(\Z+\Zc\right) \,\,,\,\, \,\,\;\;\;
\P =i\sqrt{\frac{m\w\hbar}{2}}\left(\Zc-\Z\right) \, ,
\end{equation}

\ni we get $<\X> = x \,\, , \,\, <\P> = p$, where $x$  and $p$ are
defined in the same way,
constituting the phase-space coordinates for Anti-deSitter space-time
(see
\cite{oscilata} where an adequate choice of time is discussed). We
observe
that these expectation values follow the classical trajectories
(geodesics) of the motion.

Repeating the group quantization in the new variables we obtain the
manifestly-covariant {\bf x}-representation. The states $|x,t>$ are
defined as:
\begin{equation}
|x,t> \equiv \sum_{n=0}^\infty \Psi_n^N(x,t)^* |N,n> \,,
\end{equation}

\ni where
\begin{eqnarray}
\Psi^{N}_{n}(x,t) &\equiv& e^{-in\omega t}\Phi_{n}^{N}(x)
\label{Psixt}\\
\Phi_{n}^{N}(x) &=& \sqrt{\frac{\w }{2\pi}}\left(\frac{m\w }
    {\hbar\pi}\right)^{\frac{1}{4}}\frac{1}{2^{\frac{n}{2}}\sqrt{n!}}
    \sqrt{\frac{(2N)^n}{(2N)_n}}\sqrt{\frac{\Gamma(N)}{\sqrt{N}
    \Gamma(N-\frac{1}{2})}}\alpha^{-(N+n)}H^{N}_{n}(\chi)  \, ,
\end{eqnarray}

\noindent $\alpha\equiv\sqrt{1+\frac{\omega^{2}}{c^{2}}x^{2}}$,
 $\chi\equiv\sqrt{\frac{m\omega}{\hbar}}x$ and $H^{N}_{n}(\chi)$ are
the
Relativistic Hermite polynomials \cite{RelHermite,oscilata}. These
states are
not eigenstates of the boost operator $\hat{x}$ in the same manner
that the
states $|z,t>\equiv |e^{-i\w t}z>$ are not eigenstates of the
annihilation
operator $\Z$. The integration measure is $dxdt$, coming from the
group measure
once the $p$-integration has been regularized \cite{oscilata}.

Both representations are related through the Relativistic
Bargmann transform \cite{RelBargmann}, the kernel of which is nothing
but
the configuration-space wave function of
the coherent states $|z,t>$ defined above, and have the proper
non-relativistic
limit.

The time variable can be factorized out (non-trivially) from the
manifestly-covariant {\bf x}-representation, giving rise to a {\it
minimal}
{\bf x}-representation {$|x>$}. The new integration measure turns out
to be
$dx/\alpha^2$ and it is this measure that makes the Relativistic
Hermite
polynomials (multiplied by the partial weights $\alpha^{-N-n}$) a set
of
orthogonal functions \cite{oscilata,Zarzo,Gegenbauer}.

The uncertainty relations for the operators $\X$ and $\P$ on the
$|z>$ states
are:
\begin{equation}
\Delta\X\Delta\P = \frac{\hbar}{2}\sqrt{ \kappa^2 +
    \frac{1}{4N^2}\left[ 4|z|^4-(z^2+{z^*}^2)^2 \right]} \geq
\medio\hbar\kappa=
 \medio |< \left[\X,\P\right] >|
\end{equation}

\ni The equality holds for $z=|z|e^{in\pi/2}$, i.e. $z$ real or pure
imaginary, defining the
so-called ``intelligent states" \cite{Aragone}, but only for $z=0$
(the vacuum) we reach
the absolute minimum.

Our RCS correspond to definition c) in Sec. 1, and can be identified
with the generalized coherent states on the unit complex
disk \cite{Perelomov} once the change of variables
$z_D=\sqrt{\frac{2}{N}}\frac{z}{1+\kappa} \in D\,\,(z\in C),\,\,$
and the identification $k\equiv N$
have been made, where $D$ is the unit complex disk and $k$ is the
Bargmann index
characterizing the irreducible representations of $SL(2,R)$. For a
calculation
of the uncertainty relations in the unit Disk see \cite{Eberly}.

It would be natural to ask whether a definition analogous to b) could
also be given. In fact,
there exists a solution to the relativistic eigenvalue problem,
$\Z|\rho>=\rho |\rho>$. Using the commutation relations
(\ref{commut}) and the Fock-space representation
(\ref{rep}), we obtain:
$|\rho>=c_0\sum^{\infty}_{n=0}
\frac{(\sqrt{2N}\rho)^n}{\sqrt{n!(2N)_n}}|N,n> \, ,$
where $c_0$ is an arbitrary normalization factor (a function of
$\rho$ actually). In the
\cinf limit these states also reproduce the standard non-relativistic
coherent states.
They are related to others previously defined by Barut and Girardello
\cite{Barut} through the change $\rho \rightarrow \rho/\sqrt{N}$, and
choosing
$c_0=1$. The connection between the corresponding generators is
$\Z=\frac{1}{2\sqrt{N}}\hat{L}_-$ and
$\Zc=\frac{1}{2\sqrt{N}}\hat{L}_+$.

Very recently, \cite{Trifonov}, it has been shown that both sets of
coherent
states (Perelomov's and Barut and Girardello's) are particular cases
of
a more general definition of coherent states, the generalized
intelligent
states, which minimize the Robertson-Schr\"odinger uncertainty
relation \cite{Robertson-Schrodinger}. In the particular case of
operators
satisfying canonical
commutation relations the states that minimize the
Robertson-Shr\"odinger
uncertainty relation had been called correlated states in
\cite{Manko},
although it has been proved more recently that they coincide with
ordinary
squeezed states (see \cite{Trifonov} and references therein).

\section{Canonical (higher-order) creation and annihilation
 operators: canonical, relativistic coherent states.}

The definition of polarization in group quantization (given in terms
of
left-invariant vector fields $\tilde{X}^L$) can
be generalized so as to admit operators in the left enveloping
algebra. This
generalization has already been exploited in finding a position
operator for
the free relativistic particle \cite{position}, and for obtaining a
new
momentum operator canonically asociated with the boosts operator
\cite{SinhPoincare} (as well as in solving anomalous
problems \cite{anomalia}). In the present case it also makes sense to
look for basic operators
satisfying canonical (versus manifestly covariant) commutation
relations.
Let us then seek power series in $\XLz$ and $\XLzc$ ($\eta=e^{i\w
t/2}$)
\cite{SL(2R),oscilata},
\begin{eqnarray}
\XLzc{}^{HO} &=& \XLzc +\frac{\alpha}{N}\XLz\XLz\XLzc + ... \nn \\
\tilde{X}^{L}_{\eta}{}^{HO} &=& \tilde{X}^{L}_{\eta} -\mu\XLz\XLzc
  -\frac{\nu}{N}\XLz\XLz\XLzc\XLzc + ...  \, ,
\end{eqnarray}

\ni such that $\PHO = < \tilde{X}^{L}_{\eta}{}^{HO}, \XLzc{}^{HO} >$
contains
$\tilde{X}^{L}_{\eta}$ and excludes the central generator. The
coefficients
of the power series are determined by the requirement that $\PHO$ is
a
polarization and the corresponding right operators define a unitary
action
on the wave functions $\Psi$ {\it which fortunately are the same as
before}.

\ni More specifically,
\begin{eqnarray}
\left[\tilde{X}^{L}_{\eta}{}^{HO},\XLzc{}^{HO}\right] &=& -
2\XLzc{}^{HO} \nn\\
\left[\tilde{X}^{R}_{z}{}^{HO},\tilde{X}^{R}_{z^*}{}^{HO}\right] &=&
\hat{1}
\end{eqnarray}

The resulting higher-order (canonical) creation and annihilation
operators are:
\begin{eqnarray}
\Z{}^{HO} &\equiv& \C = \Z -
\left(\frac{1}{4N}-\frac{3}{32N^2}\right)\Zc\Z\Z
+ \frac{7}{32N^2}\Zc\Zc\Z\Z\Z + ... \equiv
\sqrt{\frac{2}{1+\hat{\kappa}}}\Z \label{a(z)} \\
\Zc{}^{HO} &\equiv& \Cc =\Zc\sqrt{\frac{2}{1+\hat{\kappa}}} \nn
\end{eqnarray}

\ni and the energy operator is:
\begin{equation}
\H^{HO} = N\left(\hat{\kappa}-1\right) = \Cc\C
\end{equation}

\ni where $\hat{\kappa}\equiv \sqrt{1+\frac{2}{N}\left(\Zc\Z\right)}$
and the
operator $\sqrt{\frac{2}{1+\hat{\kappa}}}$ must be considered to be
functions of
the single operator $\left(\Zc\Z\right)$. We keep the notation
$\H^{HO}$, even
though this operator is only quadratic in $\C,\Cc$, to remind the
reader its
higher-order origin (in terms of the covariant operators $\Z,\Zc$).

The commutation relations between $\H^{HO}, \C$ and $\Cc$ are
the non-relativistic (canonical) ones, and the action of these new
operators
on the relativistic Fock states
reproduces the non-relativistic harmonic oscillator representation,
even though the states $|N,n>$ are the same relativistic energy
eigenstates as
 before.
\vskip 0.5cm

\ni {\it Canonical coherent states:}
\vskip 0.3cm

It seems quite natural to define canonical coherent states $|a>$ as
the eigenstates
of the
canonical annihilation operator, $\C|a>=a|a>$, with solutions:
\begin{equation}
|a> = e^{-|a|^2/2}\sum_n \frac{a^n}{\sqrt{n!}}|N,n> \, ,
\end{equation}

\ni and to introduce a ``non-relativistic" Bargmann-Fock space in the
usual way:
\begin{equation}
<a|N,n>=<n,N|a>^*=e^{-|a|^2/2} \frac{a^*{}^n}{\sqrt{n!}}\equiv
\tilde{\Phi}_n^{N.R.}(a) \, ,
\end{equation}

\ni with measure just $dada^*$.

\ni The connection to the relativistic Bargmann-Fock space is given
by
\begin{eqnarray}
\tilde{\Phi}_a(z)&\equiv & <z|a> = \sum_{n=0}^\infty <z|N,n><n,N|a> =
\sum_{n=0}^\infty  \tilde{\Phi}_n(z)\tilde{\Phi}_n^{N.R.}(a)^* \nn \\
  &=& \frac{1}{\pi}\sqrt{\frac{2N-1}{2N}}e^{-|a|^2/2}
   \left(\frac{1+\kappa}{2}\right)^{-N}
  \sum_{n=0}^\infty \frac{1}{n!}\sqrt{(2N)_n}
     \left(\frac{2az^*}{(2N)(1+\kappa)}\right)^n \label{<z|a>}
\end{eqnarray}

\ni This series is convergent in the entire complex plane, thus
defining
an integral function, as can be checked by standard criteria. A power
expansion in terms of $\frac{1}{N}$ can be computed:
\begin{equation}
<z|a>\approx
\frac{1}{\pi}e^{-|a|^2/2}e^{-|z|^2/2}e^{az^*}\left\{1-\frac{1}{4N}
\left[1-\medio\left(|z|^2-az^*\right)
\left(3|z|^2-az^*\right)\right]+...\right\} \nn
\end{equation}

The expectation value $<a|\Z|a>$ defines a classical function
$z=z(a)$ relating the
variables $a,a^*$ and $z,z^*$ as follows:
\begin{equation}
<a|\Z|a> = a \sum_{n=0}^\infty c_n <a|\left(\Cc\C\right)^n|a> \, ,
\end{equation}

\ni where $c_n$ are the coefficients of the power series of
$f(u)=\sqrt{1+\frac{u}{2N}}$. Then we define:
\begin{equation}
z(a)=\sqrt{1+\frac{|a|^2}{2N}}\,a \label{z(a)}
\end{equation}

Note that although $<a|\left(\Cc\C\right)^n|a> \neq
                 <a|\Cc{}^n\C{}^n|a>=|a|^{2n}$, any operator of the
form
$\hat{F}=\hat{O}\C^m$ (or $\hat{G}=\Cc{}^p\hat{O}$), where
$\left[\H^{HO},\hat{O}\right]=0$, defines a classical function $F(a)$
(or $G(a)$) by the formula:
\begin{equation}
F(a)=a^m\sum_n o_n |a|^{2n} \,\, , \,\,\,\, G(a)=a^*{}^p\sum_n o_n
|a|^{2n}\, ,
\end{equation}

\ni where $<a|\hat{O}|a>=\sum_n o_n <a|\left(\H^{HO}\right)^n|a>$ \,.

The functions
\begin{equation}
a(z)=\sqrt{\frac{2}{1+\kappa}}\,z \,\, , \,\,\;\;
a^*(z)=\sqrt{\frac{2}{1+\kappa}}\,z^*\,,
\end{equation}

\ni the inverse relation of (\ref{z(a)}), turn out to be the Darboux
coordinates taking the symplectic form
$\Omega\equiv \frac{1}{\kappa}dz\wedge dz^*$ to the canonical form
$\Omega=da\wedge da^*$.

Finally, we define
\begin{eqnarray}
\Q &\equiv& \sqrt{\frac{\hbar}{2m\w}}\left(\C+\Cc\right) \nn \\
\PI &\equiv& i\sqrt{\frac{m\w\hbar}{2}}\left(\Cc-\C\right) \,,
\end{eqnarray}

\ni satisfying
\begin{equation}
\left[\Q,\PI\right]=i\hbar\hat{1} \, ,
\end{equation}

\ni as well as their corresponding classical functions $q$ and $\pi$.
For these operators we obviously obtain
\begin{equation}
\Delta\Q\Delta\PI = \frac{\hbar}{2}
\end{equation}

\ni on the $|a>$ states.

A new minimal representation in configuration space can be introduced
which
will be called the canonical, minimal representation or
the {\bf q}-representation. The corresponding states, $|q>$, are the
eigenstates of the position operator $\Q$. They prove to be
\begin{equation}
|q> =  \left(\frac{m\w }
   {\hbar\pi}\right)^{\frac{1}{4}} \sum_{n=0}^{\infty}
 \frac{1}{2^{\frac{n}{2}}\sqrt{n!}}e^{-\xi^2/2}H_n(\xi) |N,n>
\end{equation}

\ni where $\xi\equiv \sqrt{\frac{m\w}{\hbar}}q$ and $H_n$ are the
ordinary,
non-relativistic Hermite polynomials. The integration measure is just
$dq$.

The analogue to the transformation kernel $<z|a>$ (\ref{<z|a>}) in
configuration
space, i.e. $<x|q>$, also makes sense and relates the Hermite
polynomials
and the Relativistic Hermite polynomials, and can be worked out in a
similar
way:
\begin{eqnarray}
<x|q>&=&\sum_{n=0}^\infty <x|N,n><n,N|q>=\sum_{n=0}^\infty
\Phi_n^N(x)
       \Phi_n^{N.R.}(q)^* \nn \\
 &=&\sqrt{\frac{m\w}{\hbar\pi}}e^{-\xi^2/2}\alpha^{-N}
\sqrt{\frac{\Gamma(N)}{\sqrt{N}\Gamma(N-\medio)}}\sum_{n=0}^\infty
\frac{\alpha^{-n}}{2^nn!}\sqrt{\frac{(2N)^n}{(2N)_n}}
 H_n^N(\chi)H_n(\xi)
\end{eqnarray}

\ni In the \cinf limit ($N\rightarrow\infty$) $<x|q>=\delta(x-q)$, as
corresponds
to the nonrelativistic harmonic oscillator. For large $N$ we can
compute a
power series expansion in $\frac{1}{N}$:
%
%
\begin{eqnarray}
<x|q>&\approx& \delta(x-q) +
\sqrt{\frac{m\w}{\hbar}}\frac{1}{64N}\left\{
    12(1+\chi^2)\delta(\xi-\chi) + 4\left[ 9\xi+3\chi+
\xi^3+\chi^3\right]\times
   \right. \nn \\
 & & \left. \delta'(\xi-\chi) +
   6\left[(\xi^2-\chi^2)+2\right]\delta''(\xi-\chi)
     + 4(\xi+\chi)\delta'''(\xi-\chi)\right\} + ...
\end{eqnarray}

\ni For finite $N$, we can study the convergence of the series having
into account that this depends on the large $n$ behaviour. For
$n>>N$ we can use the asymptotic expression for the relativistic
Hermite
polynomials given in \cite{Gegenbauer} (as well as the usual one for
the Hermite
polynomials) and we get (except for factors not depending on $n$)
\begin{equation}
\Phi_n^N(x) \Phi_n^{N.R.}(q)^* \sim
n^{-\frac{3}{4}}\cos\left[\sqrt{2n+1}\xi-
n\frac{\pi}{2}\right]\cos\left[(n+N)
\arcsin\frac{1}{\alpha}-N\frac{\pi}{2}\right]
\end{equation}

\ni In the particular case of $\xi=\chi=0$ the resulting series is
$\sum (2n)^{-\frac{3}{4}}$, which is of course divergent. For the
general case
the convergence of the series is assured by the convergence of the
integral
\begin{equation}
\int_\mu^\infty x^{-\frac{3}{4}}e^{i\left[ax+b\sqrt{2x+1}\right]}dx
\end{equation}

\ni with $(\mu\geq 1)$.

The existence of Galilean-like creation and annihilation operators
along with the
$SL(2,R)$ operators $\Z, \Zc$, looks rather tricky at first sight and
thus
deserves some comment. First of all, the co-existence of both type of
operators is possible only because the spectra of the
non-relativistic
and relativistic harmonic oscillator are the same and the Hamiltonian
$\hat{H}$
is shared by the two systems, although written in two different
manners:
$\hat{H}=\Cc\C=\sqrt{1+\frac{2}{N}\Zc\Z}$. The common (phase space)
Poisson algebra
contains two subalgebras $(H,a^{\dag},a)$ and $(H,z^{\dag},z)$
intersecting at H
even though $(H,a^{\dag},a,z^{\dag},z)$ does not close. The situation
is in certain aspects
similar to the case of the Schr\"odinger group
\cite{Niederer,anomalia} which (in 1+1
dimensions) is generated by an analogous set of operators with the
only difference
that the commutators between $\C,\Cc$ and $\Z,\Zc$ close and,
therefore, it is possible
to find a quantum representation in which $\Z$ and $\Zc$ are written
only as
quadratic functions of $\C$ and $\Cc$. This quantum representation is
 realized
only for the special values  of the $SL(2,R)$ Bargmann index
$k=\frac{1}{4}, \frac{3}{4}$, as a consecuence of the {\it anomalous}
character
of the Schr\"odinger group \cite{anomalia},
with direct physical application in two-photon quantum optics
\cite{Yuen}.

%

Needless to say that the non-relativistic harmonic oscillator also
support the
construction of higher-order operators $\Z, \Zc$ as functions of the
operators
$\C, \Cc$ (the inverse of (\ref{a(z)})):
$\Z=\sqrt{1+\frac{1}{2N}\H}\,\C, \Zc=
\Cc\sqrt{1+\frac{1}{2N}\H}$, thus realizing the $SL(2,R)$ group on
states ${|n>}$ and for any value of N (or Bargmann index k) and not
just for
$N=\frac{1}{4},\frac{3}{4}$ as in the case of the Schr\"odinger
group.
\vskip 0.5cm

{\bf Acknowledgement.} The authors wish to thank the referee for
valuable
suggestions.

\end{document}